\documentclass[%
 reprint,
superscriptaddress,
showpacs,
nofootinbib,
 amsmath,amssymb,
 aps,
 pra,
floatfix,c
]{revtex4-1}

\usepackage{graphicx}
\usepackage{dcolumn}
\usepackage{bm}
\usepackage{color}
\usepackage{amsmath}
\usepackage{amssymb}
\usepackage{braket}
\usepackage{float}
\usepackage[colorlinks=true,linkcolor=blue,anchorcolor=blue,citecolor=blue,urlcolor=blue]{hyperref}
\usepackage{hyperref}
\usepackage[mathlines]{lineno}

\begin{document}

\preprint{APS/123-QED}

\title{Semiclassical analysis of photoelectron interference in a synthesized two-color laser pulse}

\author{Yudi Feng}
\author{Min Li}
\email{mli@hust.edu.cn} 
\author{Siqiang Luo}
\author{Kun Liu}
\author{Baojie Du} 
\author{Yueming Zhou}
\affiliation{Wuhan National Laboratory for Optoelectronics and School of Physics, Huazhong University of Science and Technology, Wuhan 430074, China} 
\author{Peixiang Lu} 
\email{lupeixiang@hust.edu.cn}
\affiliation{Wuhan National Laboratory for Optoelectronics and School of Physics, Huazhong University of Science and Technology, Wuhan 430074, China} 
\affiliation{Hubei Key Laboratory of Optical Information and Pattern Recognition, Wuhan Institute of Technology, Wuhan 430205, China}  

\date{\today}

\begin{abstract}
We measure the photoelectron energy spectra from strong-field ionization of Kr in a two-color laser pulse consisting of a strong 400-nm field and a weak 800-nm field. The intensities of the main above-threshold ionization (ATI) and sideband peaks in the photoelectron energy spectra oscillate roughly oppositely with respect to the relative phase between the two-color components.
We study the photoelectron interferometry in strong-field ATI regime from the view of interference of different electron trajectories in order to extend RABBITT type analysis to the strong-field regime. Based on the strong-field approximation model, we obtain analytical expressions for the oscillations of both ATI and sideband peaks with the relative phase. A phase shift of $\pi/4$ with respect to the field maximum of the two-color laser pulse is revealed for the interference maximum in the main ATI peak without including the effect of the atomic potential.
\end{abstract}


\maketitle

\section{\label{sec:level1}INTRODUCTION} 
Advances in the development of laser technologies have opened the possibilities to observe the time delays in photoemission from atoms and molecules on an attosecond timescale, which provides unprecedented insight to the ultrafast electron dynamics during photoionization \cite{kruze,mairesse,Uiberacker}. Different methods have been developed to extract the temporal information of photoemissions, such as attoclock \cite{Eckle,natphy2015,HanMeng2018,GePeiPei2019}, photoelectron holography \cite{Huismans2011,Meckel,MMLiu_prl2016,YZhou_prl2016,Walt,XieHui2018,Tanjia2018,HeMingRui2018}, attosecond streak camera \cite{Itatani2002}, and RABBITT (reconstruction of attosecond harmonic beating by interference of two-photon transitions) technique \cite{Paul,Muller2002,Klunder,Vos}. In the RABBITT technique, two adjacent harmonics from an infrared (IR) pulse photoionize an atom or a molecule in the presence of the generating IR field. The interference of two-color two-photon transitions gives rise to sideband (SB) peaks between adjacent high harmonics. The SB intensity beats at twice the frequency of the IR pulse with an offset phase, from which one can extract the temporal properties of the photoemission process. Nowadays, the RABBITT interferometry has been widely used to reconstruct the relative phase of adjacent high harmonics \cite{Paul,Muller2002}, to measure the relative time delays for electron emissions from different atomic levels \cite{Klunder}, and to obtain orientation- and energy-resolved Wigner time delay in molecules \cite{Vos}. 

Recently, the RABBITT interferometry has been extended to the strong-field above-threshold ionization (ATI) regime using intense two-color laser fields (typically, 800 nm and 400 nm) \cite{Zipp}. In the strong-field ATI regime, a relatively weak field at the half frequency of a strong ionizing pulse is used to generate SB peaks between adjacent ATI peaks. The intensity of both ATI and SB peaks oscillates with respect to the relative phase between the two-color components. The phase of those oscillations encodes the relative phase between two neighboring ATI peaks \cite{Zipp}. By measuring the interfering signals from different resonant states in an orthogonally polarized two-color laser field, the ionization time delay in Freeman resonance has been observed \cite{Gong}. Based on a semiclassical model, the time delay of a temporary retrapping of a photoelectron by the atomic potential is revealed for the near threshold photoelectron \cite{Song}. Extending the photoelectron interferometry in strong-field ATI regime to chiral molecules, an attosecond time delay was revealed between electrons ejected forward and backward relative to the laser propagation direction \cite{Beaulieu}. 

Compared with the conventional RABBITT interferometry, the photoelectron interferometry in strong-field ATI regime involves multiphoton transitions, thus its mechanism is more complex. Up to now, the  photoelectron interferometry in strong-field ATI regime is interpreted in the framework of multiphoton ionization. However, in strong-field ionization, the spectral features in the photoelectron momentum distribution (PMD) are usually analyzed in terms of electron trajectories, which can give a more intuitive picture for the photoemission process. In this paper, we study the  photoelectron interferometry in strong-field ATI regime from the view of interference of different electron trajectories. We measure the photoelectron energy spectra in a strong 400-nm field combined with a weak 800-nm field with parallel polarizations. Consistent with previous studies \cite{Zipp}, the intensity of the main ATI and SB peaks oscillates with the relative phase of the two-color laser components roughly 
oppositely. By deriving analytical expressions based on the strong-field approximation (SFA), we show that those oscillations in the photoelectron energy spectra originate from the superposition of the intercycle and intracycle interferences of the released electron wave packets. An intrinsic phase shift of $\pi/4$ with respect to the field maximum is found for the interference maximum in the main ATI peak without including the effect of the atomic potential. 

Atomic units ($\hbar =\vert e\vert=m_e=1$) are used throughout this paper unless specified otherwise.

\section{\label{sec:level2} METHODS}

\subsection{Experimental method}
Experimentally,  we measure the PMDs from strong-field ionization of Kr atoms using a cold target recoil-ion momentum spectroscopy (COLTRIMS) \cite{Ullrich}. The 800-nm laser pulse is generated from an amplified Ti:sapphire femtosecond laser system. Its frequency is doubled by a 300-$\mu$m-thick $\beta$-barium-borate crystal. The intensity of the generated 400-nm laser pulse is about $5.8\times 10^{13}$ W/{cm}$^2$. The intensity ratio of the 800-nm and 400-nm fields is controlled by a dual waveplate ($\lambda /2$ for 800 nm and $\lambda$ for 400 nm) before a wire grid polarizer, which is set to be about 1:100 in our experiment. Meanwhile, a wire grid polarizer is used to ensure that the polarization directions of the 800-nm and 400-nm laser pulses are the same. The relative phase between the two color components is finely adjusted by a pair of glass wedges, which is then calibrated by comparison with the numerical calculation, as shown in Fig.\,1. Both 800- and 400-nm laser components are focused in the main chamber of the COLTRIMS by a parabolic mirror (\textit f=75 mm) and then interact with the supersonic Kr gas beam. The produced electrons and ions are guided by a uniformed electric field (about 8.8 V/cm) and a uniformed magnetic field (about 7.2 G) to the multi-channel-plate detector. Then we reconstruct the three-dimensional PMDs from the time of flights and the positions of the particles on the detectors. The PMDs are integrated with $p_z>0$ ($p_z$ is the electron momentum along the laser polarization direction) to obtain the photoelectron energy spectra with respect of  the relative phase $\phi$ between the two color components.

\subsection{SFA simulation}

The SFA method \cite{Keldysh, Faisal, Reiss} is used to study the photoelectron interference in the synthesized two-color laser fields, in which the Coulomb potential is neglected. In the SFA, the emitted electron wave packet is approximated by a plane wave. Thus, the amplitude of transition probability from the bound state $\Psi_0$ to a continuum state $\Psi_{\textbf p}$ with asymptotic momentum \textbf p can be expressed as \cite{Milosevic,LiMin2016,Kunlong}
\begin{equation}
\begin{split}
M&=-i\int_{-\infty}^{\infty}dt\langle \Psi _{\textbf p}^V(t)\mid \textbf r \cdot \textbf E(t)\mid \Psi _0\rangle \\
&=-i\int_{-\infty}^{\infty}dt\ \text e^{iS(t)}\langle \textbf p+\textbf A(t)\mid \textbf r\cdot \textbf E(t)\mid \Psi _0\rangle.
\end{split}
\end{equation}
Here $\Psi _{\textbf p}^V(t)$ is the Volkov state, which is expressed in length gauge as, 
\begin{equation}
\mid \Psi _{\textbf p}^V(t)\rangle= \mid \textbf p+\textbf A(t)\rangle \text e^{\frac i2\int^td\tau \ [\textbf p+\textbf A(\tau)]^2},
\end{equation}
and $S(t)$ is the action during the transition process,
\begin{equation}
S(t)=-\int^{\infty}_td\tau \left\{\frac 12[\textbf p+\textbf A(\tau)]^2+I_p\right\},
\end{equation}
where $I_p$ is the ionization potential, which is set to be 0.515 a.u. for Kr. Using the saddle-point approximation, the transition amplitude in Eq.\,(1) can be approximately calculated by 
 \begin{equation}
 M\backsim \sum_{t_s} \text {exp} [iS(t_s)],
 \end{equation}
 where the pre-exponential factor is omitted. $t_s$ is the complex saddle point time, which can be obtained by solving the following saddle point equation, 
\begin{equation}
\frac 1 2 [\textbf p+\textbf A(t_s)]^2+I_p=0 . 
\end{equation}

\begin{figure}
	\centering\includegraphics[width=9cm]{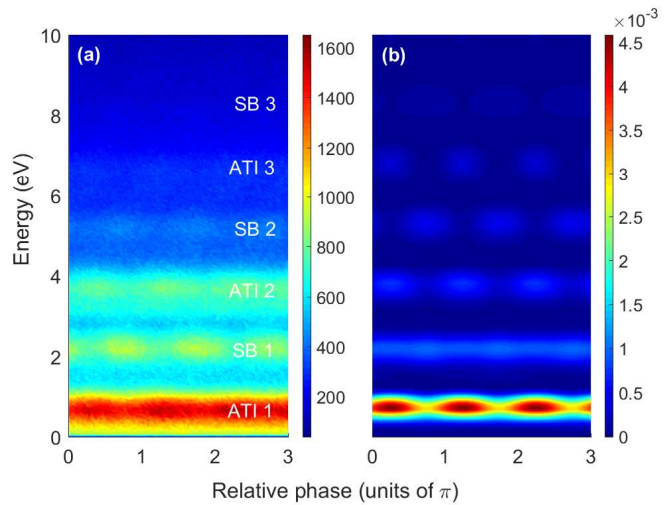}
	\caption{The photoelectron energy distributions from strong-field ionization of Kr in a two-color laser pulse with respect to the relative phase between the two-color components from the measurement (a) and from the SFA simulation (b). The laser intensity for the 400-nm field is $\sim 5.8\times10^{13} $W/cm$^2$ and for the 800-nm field is $\sim 5.8\times10^{11} $W/cm$^2$.  }
\end{figure}

The vector potential $\textbf A(t)$ of the synthesized two-color laser fields is given by,
\begin{equation}
\textbf A(t)=f(t)[-\frac {E_0} \omega \textrm{sin}(\omega t)-\frac {2\epsilon E_0} {\omega} \textrm{sin}(\frac{\omega t}{2}+\phi)]\textbf{e}_z,
\end{equation}
where $E_0$ is the electric field amplitude for the 400-nm field (The intensity for the 400-nm field is $5.8\times10^{13}$W/cm$^2$ and for the 800-nm field is $5.8\times10^{11}$W/cm$^2$ regulated by $\epsilon$),  $\omega$ is the angular frequency of the 400-nm laser pulse, and the pulse envelope $f(t)=\textrm{sin}^2(\pi t/T_p)$ is employed with a duration of $T_p=16T_{400}$, where $T_{400}$ is the period of the 400-nm laser field, $\phi$ is the relative phase between the two-color components, $\textbf {e}_z$ is the unit vector along the laser polarization direction. We calculate the PMDs and the photoelectron energy spectra at different relative phases $\phi$ and obtain the $\phi$-dependent energy spectra.

\section{\label{sec:level3}Results and discussions}

In Fig.\,1(a), we show the measured photoelectron energy spectra of Kr in synthesized 400-nm and 800-nm fields with parallel polarizations as a function of the relative phase between the two-color components. The photoelectron energy spectra are obtained with $p_z>0$. One can see that, when the perturbative 800-nm field is added, the SB peaks [labeled as SB 1, 2, 3 in Fig.\,1(a)]  emerge between adjacent 400-nm main ATI peaks [labeled as ATI 1, 2, 3 in Fig.\,1(a)]. The intensity of both main ATI and SB peaks oscillates with respect to the relative phase $\phi$. Furthermore, the intensity of the main-ATI and SB peaks oscillates roughly oppositely with respect to $\phi$, which is very similar to the photoelectron spectra in conventional RABBITT method \cite{Paul,Muller2002,Klunder,Vos}. The simulated $\phi$-dependent photoelectron energy spectra by the SFA are shown in Fig.\,1(b), which agree well with the measured results.

\begin{figure}	
	\includegraphics[width=9cm]{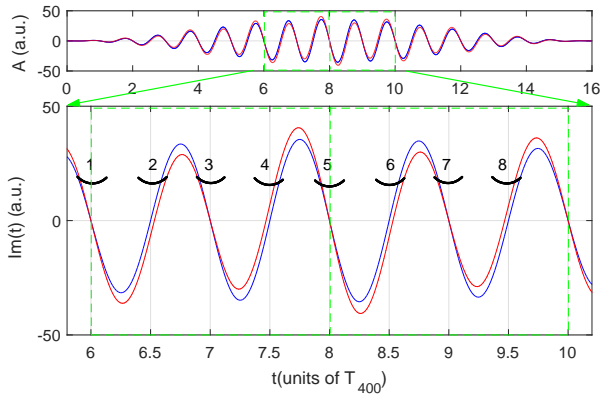}
	\caption{\label{fig2}
		(a) The vector potentials of the 400-nm laser field (blue) and the synthesized two-color laser field (red) used in the SFA simulation. (b) Eight saddle-point solutions (labeled by the numbers) for the ATI in the synthesized two-color laser pulse when the ionization time is within [6, 10]$T_{400}$. The laser vector potentials of the 400-nm field and the synthesized two-color field are shown by the blue and red curves, respectively, in arbitrary units. The relative phase is zero for the synthesized two-color laser field.
	}	
\end{figure}

Next we use the SFA to analyze the origin of the oscillations for the main ATI and SB peaks. In Fig.\,2(a), we show the vector potential of the synthesized two-color laser field used in the SFA simulation with $\phi =0$. Due to the perturbative nature of the 800-nm field, the vector potential of the synthesized two-color laser field is very close to that of the single-color 400-nm field. In Fig.\,2(b), we show eight saddle-point solutions in the middlemost part of the synthesized two-color laser pulse by the black segments, corresponding to two periods of the two-color field. Each black segment corresponds to a series of saddle point solutions with different asymptotic momenta. In fact, The SB peaks are involved with the intercycle photoelectron
interference of 800 nm field \cite{HanMeng2018}. Thus the PMDs in the two-color laser field can be generally described by the interference of those eight saddle-point solutions. 

To show which saddle-point solutions have large contributions to the oscillations in the energy spectra, we compare the standard SFA results with the results with only considering parts of the saddle points in Fig.\,3. In Fig.\,3(a), the SFA result includes all the saddle-point solutions, and it is normalized in order to improve the contrast for each main ATI or SB peak. In Fig.\,3(b),  eight saddle-point solutions [saddle points 1-8 in Fig.\,2(b)] are considered, while in Fig.\,3(c) only four saddle point solutions [saddle points 1, 3, 5 and 7 in Fig.\,2(b)] are considered. We can see that the oscillations for both main ATI and SB peaks appear when only four saddle-point solutions are considered. The intensity of the main ATI and SB peaks oscillates roughly $\pi$ out of phase, which is consistent with the measurement. Thus the oscillations of the main ATI and SB peaks observed in the experiment can be interpreted as the interference of the saddle-point solutions of 1, 3, 5, and 7. The interference pattern of the saddle-point solutions of 2, 4, 6, and 8 are very similar to that in Fig.\,3(c).

\begin{figure}
	\centering\includegraphics[width=9cm]{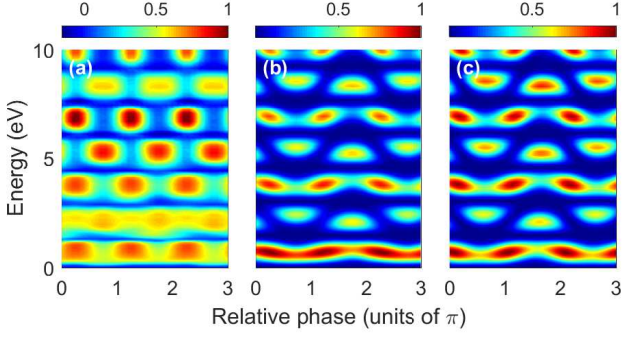}
	\caption{The photoelectron energy distributions with respect to the relative phase between the two-color components from the SFA simulations. In (a), all SP solutions are included, and the spectra are normalized to improve the contrast for each main ATI or SB peak. In (b), the SP solutions of 1-8 in Fig.\,2(b) are included. In (c), the SP solutions of 1, 3, 5, and 7 in Fig.\,2(b) are included. In (b) and (c), we only includes the integral from the real part of the saddle-point time to the end of the laser pulse for the classical action in the transition amplitude.}
\end{figure}

In the SFA simulation, the saddle-point time $t_s$ is complex. The imaginary part of $t_s$ leads to a sub-barrier phase, which has a significant effect on the interference pattern in the PMD \cite{Han}. The imaginary part of $t_s$ is also closely related to the ionization rate. To show whether the sub-barrier phase and the ionization rate have a significant effect on the oscillations of the ATI and SB peaks, in Figs.\,3(b) and 3(c), we have replaced the lower limit of the integral in Eq.\,(3) by the real part of $t_s$, i.e., the effect of the imaginary part of $t_s$ has been neglected. One can see that the oscillations are very similar to those in Fig.\,3(a) except for the first SB peak. Thus, the ionization rate and the sub-barrier phase play a minor role in forming the oscillations of the main ATI and SB peaks. 

Based on the above analysis, we can use a simplified analytical model to interpret the oscillations of the main ATI and SB peaks. By setting the pulse envelope in Eq.\,(6) to be 1, the vector potential and electric field of the synthesized two-color laser pulse can be expressed, respectively, as
\begin{align}
\textbf{A}(t)&=-[\frac {E_0} \omega \textrm{sin}(\omega t)+\frac {2\epsilon E_0} {\omega} \textrm{sin}(\frac{\omega t}{2}+\phi)]\textbf{e}_z ,\\
\textbf{E}(t)&=[E_0 \textrm{cos}(\omega t)+\epsilon E_0 \textrm{cos}(\frac{\omega t}{2}+\phi)]\textbf{e}_z.
\end{align}
Because the trajectories corresponding to those saddle-point solutions 1, 3, 5, and 7 in Fig.\,2(b) are important, the momentum distribution $M(p)$ can be simplified to,
\begin{equation}
\begin{split}
M(p)&=\left| \sum_{i=1,3,5,7}e^{iS(t_s^{(i)})}\right|^2 \\
&=\left|[(e^{iS_1}+e^{iS_3})+e^{iS_{51}}(e^{iS_1}+e^{iS_3})]\right|^2 \\
&=4[1+\cos(S_{51})][1+\cos(S_{31})],
\end{split}
\end{equation}
where $\Delta S_{ij}=S(t_s^{(i)})-S(t_s^{(j)}) (i,j=1,3,5,7)$ represents the phase difference between two trajectories. Here we have neglected the effect of the ionization rate. Since the release times of the trajectories 5 and 1 are separated by one optical cycle, $\textrm{cos}(\Delta S_{51})$ corresponds to the intercycle interference of the released electron wave packet \cite{Arbo2010,LiMin2014,LiLiang}.  Because the trajectories 3 and 1 are released within the same optical cycle, $\textrm{cos}(\Delta S_{31})$ corresponds to the intracycle interference.

We use $t_0$ to denote the ionization time of the trajectory 1 ($t_0$=0 corresponds to the field maximum). As a result, the ionization time of the trajectory 5 is $ t_0+2T_{400} $. Using the vector potential in Eq.\,(7), the phase difference between trajectories 5 and 1 can be analytically given by,
\begin{equation}
\begin{split}
\Delta S_{51}&=\int_{t_0}^{t_0+2T_{400}}d\tau \{\frac 12[p+A(\tau)]^2+I_p\},\\
&=(E_k +I_p+U_p) 2 T_{400},
\end{split}
\end{equation}
where $U_p={E_0^2}/{4\omega ^2}+{\epsilon ^2E_0^2}/{\omega ^2}$ is the ponderomotive energy of the synthesized laser fields and $E_k={p^2}/2$ is the final electron energy corresponding to the asymptotic momentum $p$.

Since the 800-nm field is weak, we can assume that the difference of the ionization time between the trajectories 3 and 1 is almost $T_{400}$. Thus the ionization time of the trajectory 3 can be approximately given by $t_0+T_{400}$. Using the vector potential in Eq.\,(7), the phase difference between trajectories 3 and 1 is expressed as, 
\begin{equation}
\begin{split}
\Delta S_{31}=&\int_{t_0}^{t_0+T_{400}}d\tau  \{\frac 12[p+A(\tau)]^2+I_p\}\\
=&\int_{t_0}^{t_0+T_{400}}d\tau  (\frac 12p^2+I_p) \\
&+\int_{t_0}^{t_0+T_{400}}d\tau[\frac 12A_{400}^2(\tau)
+\frac 12  A_{800}^2(\tau)+pA_{400}(\tau)]\\
&+\int_{t_0}^{t_0+T_{400}}d\tau  [p+A_{400}(\tau)]A_{800}(\tau)\\
=&(E_k +I_p+U_p)T_{400}+
\frac{4\epsilon E_0^2}{\omega ^3} \textrm{sin}(\phi-\frac{\omega t_0}{2})\\
&+\frac{4\epsilon E_0^2}{3\omega ^3} \textrm{sin}(\phi+\frac{3\omega t_0}{2})
-\frac{8\epsilon pE_0}{\omega ^2} \textrm{cos}(\phi+\frac{\omega t_0}{2}),
\end{split}
\end{equation}
where $ A_{400}(t) $ and $ A_{800}(t) $ are the vector potentials of the 400-nm and 800-nm laser fields, respectively.

Because the electron wave packets are released near the maximum of the laser field, $ wt_0 $ is small as compared with $ \phi $. Therefore, the dependence of the phase difference $\Delta S_{31}$ on the ionization time can be omitted, i.e., $\Delta S_{31}$ can be approximately given by,
\begin{equation}
\begin{split}
\Delta S_{31}
\approx &(E_k +I_p+U_p)T_{400}+ \\
&\frac{16\epsilon E_0^2}{3\omega ^3} \textrm{sin}\phi
-\frac{8\epsilon pE_0}{\omega ^2} \textrm{cos}\phi.
\end{split}
\end{equation}
This phase difference can be rewritten as,
\begin{equation}
\Delta S_{31}=(E_k +I_p+U_p)T_{400}+\alpha \textrm{sin}(\phi+\phi_0),
\end{equation}
where $\alpha=\frac {8\epsilon E_0}{3\omega ^3} \sqrt{9 p^2\omega ^2+4E_0^2}$ is a scaling factor, and $\phi_0 =\textrm {arctan} \left(-\frac {3\omega p}{2E_0} \right)$ is a phase shift. Substituting Eqs.\,(10) and \,(13) into Eq.\,(9), we obtain analytical expressions for the $\phi$-dependent photoelectron energy spectra.

\begin{figure}
	\centering\includegraphics[width=9cm]{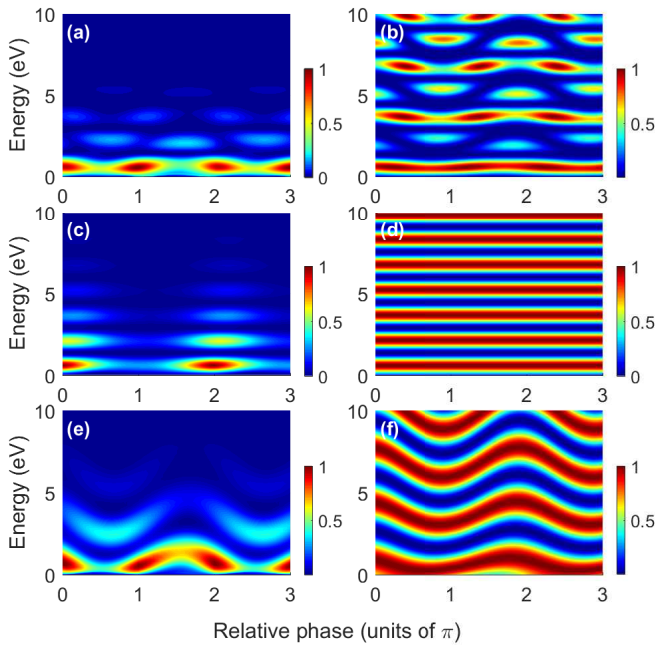}
	\caption{The photoelectron energy distributions with respect to the relative phase $\phi$ from the SFA simulations (left column) and the analytical model (right column). (c) and (d) show the intercycle interferences of the trajectories 5 and 1, (e) and (f) show the intracycle interferences of the trajectories 3 and 1, and (a) and (b) show the superposition of the intercycle and intracycle interferences, i.e., the interferences of the trajectories of 7, 5, 3, and 1. }
\end{figure}

To validate the analytical model, we compare the results by the analytical model with the SFA in Fig.\,4. Figure 4(c) shows the intercycle interference of the trajectories 5 and 1 calculated by the SFA. One can see that the intercycle interference reveals the ATI-like structures along the electron energy direction. The intercycle interference calculated by the analytical model, i.e., $\textrm{cos}(\Delta S_{51})$, agrees well with the SFA, as shown in Fig.\,4(d). According to Eq.\,(10), the phase difference $\Delta S_{51}$ is not a function of the relative phase $\phi$, thus the intercycle interference fringes show no dependence on the relative phase. Figure 4(e) shows the intracycle interference of the trajectories 3 and 1 calculated by the SFA, which reveals a wavelike interference pattern. This wavelike pattern is also reproduced by the analytical model [$\textrm{cos}(\Delta S_{31})$], as shown in Fig.\,4((f). One can also see that the fringe spacing of the intracycle interference in Fig.\,4(f) is twice as large as that of the intercycle interference in Fig.\,4(d). The interference patterns including the intercycle and intracycle interferences by the analytical model are shown in Fig.\,4(b), which is also consistent with the SFA [Fig.\,4(a)]. 

Using the analytical model, we can simply explain the reason why the main ATI and SB peaks in the energy spectra oscillate roughly $\pi$ out of phase with respect to $\phi$. For the main ATI peak, the photoelectron is released with absorbing an integer number ($n$) of the 400-nm photons, or equivalently, an even number ($2n$) of the 800-nm photons. Thus the electron energy satisfies $ E_k=2n\omega _{800} -I_p-U_p$, where $\omega_{800}=w/2$ is the frequency of the 800-nm laser. For the SB peaks corresponding to an absorption of an odd number of the 800-nm photons, the electron energy satisfies $ E_k=(2n+1)\omega _{800} -I_p-U_p$. According to Eq.\,(10), we know that the main ATI and SB peaks correspond to the maxima of the intercycle interferences. Substituting those two equations into Eq.\,(13), the intracycle interference of the trajectories 3 and 1 for the main ATI and SB peaks becomes,
\begin{equation}
\textrm{cos}(S_{31})=
\begin{cases}
 \textrm{cos}[2n\pi +\alpha \textrm{sin}(\phi+\phi_0)]& \text{main ATI}\\
\textrm{cos}[(2n+1)\pi +\alpha \textrm{sin}(\phi+\phi_0)]& \text{sideband}
\end{cases}
\end{equation}
Thus the intensities of the main ATI and SB peaks oscillate $\pi$ out of phase with respect to the relative phase $\phi$. 

The principal reason of the opposite oscillations for the main ATI and SB peaks can be explained simply. In the two-color laser field, the emission time events for the intercycle interference are spaced by $T$ while for the intracycle interference are spaced by almost $T/2$, where $T$ is the period of the two-color laser field. Therefore, the intercycle interference results in a spectrum with fringe spacing of one 800-nm photon energy, while the intracycle interference results in a spectrum with fringe spacing of twice the 800-nm photon energy, as shown in Figs.\,4(d) and 4(f). When the intracycle interference is superimposed on the intercycle interference, every other intercycle interference maximum is eliminated by destructive intracycle interference depending on the relative phase of the two-color field. Therefore the intensities of the main ATI and the SB peaks oscillate $\pi$ out of phase.

The phase of the oscillation for the main ATI and SB peaks is essential for extracting the temporal property of the photoelectron emission process. Next we concentrate on the absolute phase of the oscillation for the main ATI. For the laser field given by Eqs.\,(7) and (8), it might be expected that the maximum of the oscillation appears at the phase of zero since the field maxima of the two-color laser components coincide at $\phi=0$. However, one can see a clear phase shift of $\sim \pi/4$ with respect to the field maximum for the oscillation of the main ATI peak. This phase shift is nearly independent on the electron energy.

\begin{figure}	
	\includegraphics[width=10cm]{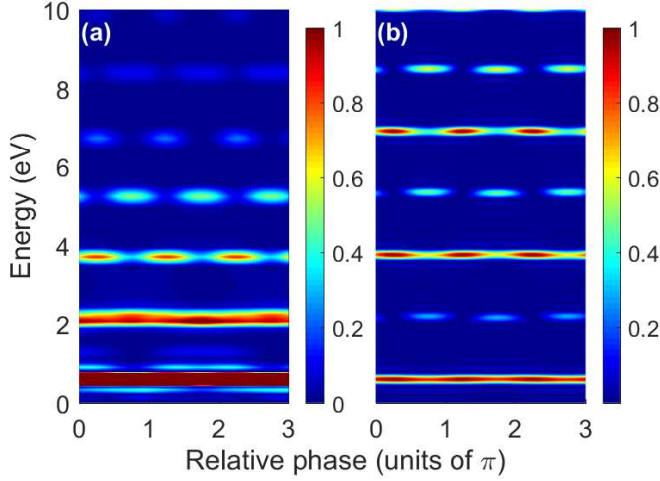}
	\caption{
		The photoelectron energy spectra with respect to the relative phase $\phi$ calculated by the SFA (a) and by the analytical model [Eq.\,(17)] (b). In the SFA simulation, the pulse envelope is set to be 1. In (b), a factor of the intercycle interference is multiplied to give rise to the ATI-like peaks \cite{Arbo2010}. 
	}	
\end{figure}

We use the analytical model to show the origin of the intrinsic phase shift. In fact, there is a phase shift $\phi_0$ predicted by Eq.\,(13). However, the phase shift $\phi_0$ depends on the electron energy, which disagrees with the SFA simulation. The main reason of the difference is that we only include two saddle-point solutions within a laser cycle in Eq.\,(13). In Fig.\,2(b), one can see that there are four saddle-point solutions within a laser cycle. To consider the contributions of the saddle-point solutions of 4 and 2 to the photoelectron momentum distributions, we obtain the phase difference between the trajectories 4 and 2,
\begin{equation}
\begin{split}
\Delta S_{42}\approx&(E_k +I_p+U_p) T_{400}\\
&+\frac{8\epsilon pE_0}{w^2}\sin(\phi)-\frac{16\epsilon E_0^2}{3w^3}\cos(\phi).\\
\end{split}
\end{equation}
Thus, the intracycle interference of the trajectories 1, 2, 3, and 4 can be expressed as (with neglecting the intra-half-cycle interference, e.g., the interference between trajectories 1 and 2),
\begin{equation}
\begin{split}
M_{\textrm{intra}}&=\textrm{cos}^2(\frac{\Delta S_{31}}2)+\textrm{cos}^2(\frac{\Delta S_{42}}2)\\
	&=1+\textrm{cos}(\frac{\Delta S_{31}+\Delta S_{42}}2)
	\textrm{cos}(\frac{\Delta S_{31}-\Delta S_{42}}2)\\
	&\approx 1+\textrm{cos}(\frac{\Delta S_{31}+\Delta S_{42}}2).\\
\end{split}
\end{equation}
Here we have assumed that the difference between $\Delta S_{31}$ and $\Delta S_{42}$ is very small. Subtituting Eqs.\,(12) and (15) into Eq.\,(16), we obtain,
\begin{equation}
\begin{split}
M_{\textrm{intra}}&=1+\cos\left[\frac{\Delta S_{51}}{2}+{\frac{\sqrt 2}2}(A+B)\sin (\phi-\frac \pi 4) \right].
\end{split}
\end{equation}
where $ \Delta S_{51}$ is given by Eq.\,(10), $ A=16\epsilon E_0^2/(3w^3) $, and $ B=8\epsilon pE_0/w^2 $. For the main ATI peaks ($ E_k=2n\omega _{800} -I_p-U_p$), equation (17) becomes,
\begin{equation}
M_{\textrm{intra}}=1+\cos[{\frac{\sqrt 2}2}(A+B)\sin (\phi-\frac \pi 4)].
\end{equation}
Thus we obtain the intrinsic phase shift of $\pi/4$ for the oscillation of the main ATI peak, which agrees well with the SFA simulation. In Fig.\,5, we compare the SFA with the analytical result by Eq.\,(17). Here the pulse envelope is set to be 1 in the SFA simulation. One can see that the oscillations for both main ATI and SB peaks by the analytical model agree with the SFA simulation. In the analytical model, the phase shift of $\pi/4$ comes from the last two terms of Eqs.\,(12) and (15). According to Eq.\,(11), we know that those two terms originate from the integral of $\int_{t_0}^{t_0+T_{400}}d\tau  [p+A_{400}(\tau)]A_{800}(\tau)$. Thus the intrinsic phase shift comes from the effect of the weak 800-nm field on the electron trajectory phase. 

\begin{figure}
	\centering\includegraphics[width=9cm]{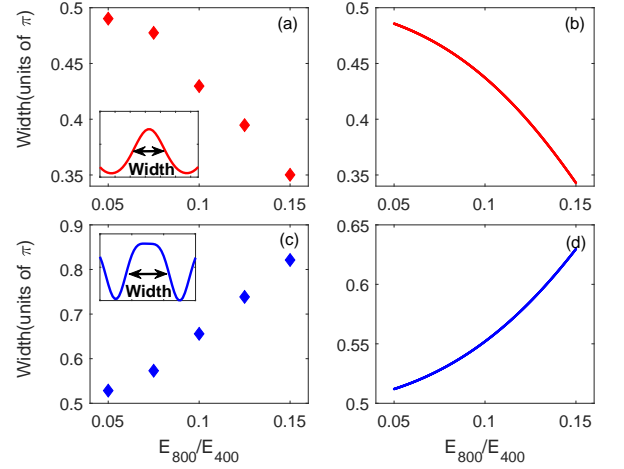}
	\caption{The width of the oscillation as a function of the field strength ratio between the 800-nm and 400-nm fields for the main ATI peak [(a) and (b)] and the SB peak [(c) and (d)]. The left and right columns correspond to the results from the SFA simulation and the analytical model, respectively. The insets in (a) and (c) show the lineouts taken from Fig.\,5(a) for the main ATI and SB peaks, respectively. The arrows show the width of the oscillation with respect to the relative phase.}
\end{figure}

It should be noted that the oscillations of the main ATI and SB peaks with the relative phase $\phi$ in the analytical model have the form of $\cos[\alpha \sin(\phi+\phi_0)]$, which is different from the simple sinusoidal modulation used in some previous studies \cite{Zipp,Gong,Song}. According to our calculation, the sinusoidal modulation is only valid when the scaling factor $\alpha$ is small, which corresponds to a very weak 800-nm field. When the scaling factor $\alpha$ is not very small, the oscillations of the main ATI and SB peaks will deviate from the sinusoidal modulation, as shown in the insets of Figs.\,6(a) and 6(c). We use the width of the oscillation (full width at half maximum) to indicate the degree of deviation from the sinusoidal modulation. In Figs.\,6(a) and 6(c), one can see that the oscillation is similar to a sinusoidal modulation when the field strength ratio $\epsilon$ is smaller than 0.05, corresponding to a width of $\sim 0.5\pi$. With increasing the field strength ratio, the width of the oscillation for the main ATI peak decreases while for the SB peak increases. This means that the oscillations for both main ATI and SB peaks deviate from the sinusoidal modulation when the field strength ratio is comparably large. The prediction of the analytical model shows the same tendency as the SFA simulation, as shown in Figs.\,6(b) and 6(d). 

\begin{figure}
	\centering\includegraphics[width=9cm]{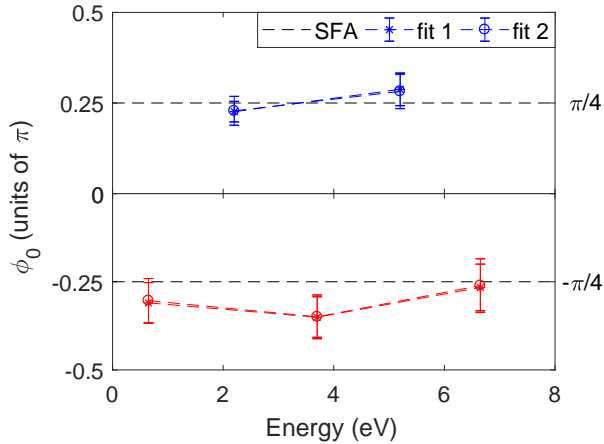}
	\caption{The phase $\phi_0$ of the oscillation for the main ATI (red) and SB (blue) peaks obtained by fitting the experiment data in Fig.\,1(a). The stars are obtained using the fitting fuction of ``$\cos(x+\phi_0)$" (fit 1) and the circles are obtained using the fitting function of ``$\cos[\alpha \sin(x+\phi_0)]$" (fit 2). The SFA results are shown by the black dash lines.}
\end{figure}

Finally, we discuss the effect of the atomic potential on the oscillations of the main ATI and SB peaks. We use two fitting functions [$\cos(x+\phi_0)$ and $\cos[\alpha \sin(x+\phi_0)]$] to obtain the phase $\phi_0$ of each main ATI or SB peak from the experiment data in Fig.\,1(a). The results are shown in Fig.\,7. The stars are obtained using the fitting fuction of ``$\cos(x+\phi_0)$" (fit 1) and the circles are obtained using the fitting function of ``$\cos[\alpha \sin(x+\phi_0)]$" (fit 2). One can see that the results using those two fitting functions are nearly the same, which implies that the oscillations of the main ATI and SB peaks can be approximated by the simple sinusoidal modulation in our experiment condition. The fitted $\phi_0$ from the measurement deviates from the SFA prediction for both main ATI and SB peaks. This small deviation originates from the effect of the Coulomb potential on the electron trajectory phase, which is neglected in the SFA simulation. This Coulomb effect can be analyzed using a Coulomb-corrected semiclassical model \cite{Song}. Moreover, it has recently shown that the RABBITT-like photoelectron interferometry in the strong-field ATI regime can be used to measure the relative attosecond delays induced by Freeman resonance \cite{Gong} and to reveal the resonant photoionization dynamics in chiral molecules \cite{Beaulieu}. Those intermediate resonant Rydberg states are not included in the SFA, thus the resonance is usually interpreted within the picture of multiphoton ionization. The role of the resonance in the strong-field photoionization dynamics might also be intuitively understood from the view of interference if the effect of the Rydberg state could be included in the SFA \cite{HuShilin}.

\section{\label{sec:level4}Conclusion}
In summary, we have measured the photoelectron energy spectra in a synthesized two-color laser pulse. The main ATI and SB peaks observed in the energy spectra oscillates $\pi$ out of phase with respect to the relative phase between the two-color components, which is consistent with previous studies \cite{Zipp}. Using the SFA method, we have systematically studied those oscillations from the view of interference of different electron trajectories. We show that the fringe spacing of the intracycle interference pattern is twice of the intercycle interference pattern. Thus every other intercycle interference maximum is eliminated by destructive intracycle interference, leading to opposite oscillations for the main ATI and SB peaks. Analytical expressions are obtained based on the SFA. We show that those oscillations for the main ATI and SB peaks have the form of ``$\cos[\alpha \sin(x+\phi_0)]$" , which deviates from the simple sinusoidal modulation when the field strength ratio between the two-color components is comparably large. Moreover, an intrinsic phase shift of $\pi/4$ with respect to the laser field maximum has been found for the interference maximum without including the effect of the atomic potential on the electron, which comes from the effect of the weak 800-nm field on the electron trajectory phase. Our work provides an intuitive picture for the study of attosecond time delays in photoemissions from atoms and molecules in the strong-field ATI regime, which is also significant for the interpretation of the Coulomb effect on the emission of the near-threshold photoelectrons \cite{Zipp,Song}.

\section{acknowledgment}

This work is supported by the National Natural Science Foundation of China (Grants No. 11722432, No. 11674116, and No. 61475055) and Program for HUST Academic Frontier Youth Team. 


\begin{thebibliography}{99}

\bibitem{kruze}
	F. Krausz and M. Ivanov, Rev. Mod. Phys. \textbf{81}, 163 (2009).

	\bibitem{mairesse}
	Y. Mairesse \textit{et al.}, Science {\bf 302}, 1540 (2003).

	\bibitem{Uiberacker}
	M. Uiberacker \textit{et al.}, Nature (London) {\bf 446}, 627 (2007).
	
	\bibitem{Eckle}
	P. Eckle, A. N. Pfeiffer, C. Cirelli, A. Staudte, R. D\"{o}rner, H. G. Muller, M. B\"{u}ttiker, and U. Keller, Science \textbf{322}, 1525 (2008).
	
	\bibitem{natphy2015}
	Torlina \textit{et al.}, Nat. Phys. \textbf{11}, 503 (2015).
			
	\bibitem{HanMeng2018}
	M. Han, P. Ge, Y. Shao, Q. Gong, and Y. Liu, Phys. Rev. Lett. \textbf{120}, 073202 (2018).
	
	\bibitem{GePeiPei2019}
	P. Ge,  M. Han, Y. Deng, Q. Gong, and Y. Liu, Phys. Rev. Lett. \textbf{122}, 013201 (2019).

	\bibitem{Huismans2011}
	Y. Huismans \textit{et al.}, Science \textbf{331} 61 (2011).
	
	\bibitem{Meckel}
	M. Meckel, A. Staudte, S. Patchkovskii, D. M. Villeneuve, P. B. Corkum, R. D\"{o}rner, and M. Spanner, Nat. Phys. \textbf{10}, 594 (2014).
	\bibitem{MMLiu_prl2016}
	M.-M. Liu, M. Li, C. Wu, Q. Gong, A. Staudte, and Y. Liu, Phys. Rev. Lett. \textbf{116}, 163004 (2016).
	\bibitem{YZhou_prl2016}
	Y. Zhou, O. I. Tolstikhin, and T. Morishita, Phys. Rev. Lett. \textbf{116}, 173001 (2016).
	
	\bibitem{Walt}
	S. G. Walt, N. B. Ram, M. Atala, N. I. Shvetsov-Shilovski, A. von Conta, D. Baykusheva, M. Lein, and H. J. W\"{o}rner, Nat. Commun. \textbf{8}, 15651 (2017).
	
	\bibitem{HeMingRui2018}
	M. He, Y. Li, Y. Zhou, M. Li, W. Cao, and P. Lu, Phys. Rev. Lett. \textbf{120}, 133204 (2018).

	\bibitem{XieHui2018}
	M. Li \emph{et al.}, Phys. Rev. Lett. \textbf{122}, 183202 (2019); H. Xie, M. Li, S. Luo, Y. Li, J. Tan, Y. Zhou, W. Cao, P. Lu, Opt. Lett. \textbf{43}, 3220 (2018).

	\bibitem{Tanjia2018}
	J. Tan, Y. Zhou, M. He, Y. Chen, Q. Ke, J. Liang, X. Zhu, M. Li, and P. Lu, Phys. Rev. Lett. \textbf{121}, 253203 (2018); J. Tan, Y. Zhou, M. He, Q. Ke, J. Liang, Y. Li, M. Li, and P. Lu, Phys. Rev. A \textbf{99}, 033402 (2019); Q. Ke, Y. Zhou, J. Tan, M. He, J. Liang, Y. Zhao, M. Li, and P. Lu, Opt. Express \textbf{27}, 32193 (2019).
	
	\bibitem{Itatani2002}
	J. Itatani, F. Qu\'er\'e, G. L. Yudin, M. Yu. Ivanov, F. Krausz, and P. B. Corkum, Phys. Rev. Lett. \textbf{88}, 173903 (2002).
	
	\bibitem{Paul}
	P. M. Paul, E.S. Toma, P. Breger, G. Mullot, F. Aug\'{e}, P. Balcou, H. G. Muller, and P. Agostini, Science \textbf{292}, 1689 (2001).
	
	\bibitem{Muller2002}
	H. G. Muller, Appl. Phys. B, {\bf 74}, s17 (2002).
	
	\bibitem{Klunder}
	K. Kl\"under, J. M. Dahlstr\"om, M. Gisselbrecht, T. Fordell, M. Swoboda, D. Gu\'enot, P. Johnsson, J. Caillat, J. Mauritsson, A. Maquet, R. Ta\"ieb, and A. L'Huillier, Phys. Rev. Lett. \textbf{106}, 143002 (2011).

	\bibitem{Vos}
	J. Vos, L. Cattaneo, S. Patchkovskii, T. Zimmermann, C. Cirelli, M. Lucchini, A. Kheifets, A. S. Landsman, U. Keller, Science \textbf{360}, 1326 (2018).

    \bibitem{Zipp}
   	L. Zipp, A. Natan, and P. Bucksbaum, Optica \textbf{1}, 361 (2014).

	\bibitem{Gong}
	X. Gong, C. Lin, F. He, Q. Song, K. Lin, Q. Ji, W. Zhang, J. Ma, P. Lu, Y. Liu, H. Zeng, W. Yang, and J. Wu, Phys. Rev. Lett. \textbf{118}, 143203 (2017).
	
	\bibitem{Song}
	X. Song, G. Shi, G. Zhang, J. Xu, C. Lin, J. Chen, and W. Yang, Phys. Rev. Lett. \textbf{121}, 103201 (2018).
		
	\bibitem{Beaulieu}
	S. Beaulieu \textit{et al.}, Science \textbf{358}, 1288 (2017).
	
	\bibitem{Ullrich}
	J. Ullrich, R. Moshammer, A. Dorn, R. D\"orner, L. Ph. H. Schmidt, and H. Schmidt-B\"ocking, Rep. Prog. Phys. \textbf{66}, 1463 (2003).
	
	\bibitem{Keldysh}
	L. V. Keldysh, Sov. Phys. JETP \textbf{20}, 1307 (1965).
	
	\bibitem{Faisal}
	F. H. M. Faisal, J. Phys. B \textbf{6}, L89 (1973).
	
	\bibitem{Reiss}
	H. R. Reiss, Phys. Rev. A \textbf{22}, 1786 (1980).

	\bibitem{Milosevic}
	D. B. Milo\v{s}evi\'{c}, G. G. Paulus, D. Bauer, and W. Becker, J. Phys. B \textbf{39}, R203 (2006).
			
	\bibitem{LiMin2016}
	M. Li, J.-W. Geng, M. Han, M.-M. Liu, L.-Y. Peng, Q. Gong, and Y. Liu, Phys. Rev. A \textbf{93}, 013402 (2016); M. Li, M.-M. Liu, J.-W. Geng, M. Han, X. Sun, Y. Shao, Y. Deng, C. Wu, L.-Y. Peng, Q. Gong, and Y. Liu, Phys. Rev. A \textbf{95}, 053425 (2017). 
	
	\bibitem{Kunlong}
	K. Liu, S. Luo, M. Li, Y. Li, Y. Feng, B. Du, Y. Zhou, P. Lu, and I. Barth, Phys. Rev. Lett. \textbf{122}, 053202 (2019); S. Luo, M. Li, W. Xie, K. Liu, Y. Feng, B. Du, Y. Zhou, and P. Lu, Phys. Rev. A \textbf{99}, 053422 (2019); W. Xie, M. Li, S. Luo, M. He, K. Liu, Q. Zhang, Y. Zhou, and P. Lu. Phys. Rev. A \textbf{100}, 023414 (2019); Y. Zhao, Y. Zhou, J. Liang, Z. Zeng, Q. Ke, Y. Liu, M. Li, and P. Lu, Opt. Express \textbf{27}, 21689 (2019).
	
	\bibitem{Han}
	M. Han, P. Ge, Y. Shao, M.-M. Liu, Y. Deng, C. Wu, Q. Gong, and Y. Liu, Phys. Rev. Lett. \textbf{119}, 073201 (2017).
	
	\bibitem{Arbo2010}
	D. G. Arb\'{o}, K. L. Ishikawa, K. Schiessl, E. Persson, and J. Burgd\"{o}rfer, Phys. Rev. A \textbf{82}, 043426 (2010).
	
	\bibitem{LiMin2014}
	M. Li, J.-W. Geng, Hong Liu, Y. Deng, C. Wu, L.-Y. Peng, Q. Gong, and Y. Liu, Phys. Rev. Lett. {\bf 112}, 113002 (2014).
	
	\bibitem{LiLiang}
	B. Wang, L. He, Y. Qing, Y. Zhang, R. Shao, P. Lan, and P. Lu, Opt. Express {\bf 27}, 30172 (2019); L. Li, P. Lan, L. He, W. Cao, Q. Zhang, and P. Lu, arXiv:1908.07283
	
	\bibitem{HuShilin}
	S. Hu \textit{et al.}, Opt. Express \textbf{27}, 31629 (2019).




					


\end{thebibliography}

\end{document}